\documentclass[a4paper,notitlepage, 12pt]{article}
\usepackage{amsfonts}
\usepackage{amssymb}
\usepackage{amsmath}
\usepackage{bm}
\usepackage{graphicx}
\def\one{{\hbox{1\kern-.8mm l}}}

\usepackage{color}
\usepackage{graphics}

\newcommand{\beq}{\begin{equation}}
\newcommand{\eeq}{\end{equation}}
\newcommand{\be}{\begin{eqnarray}}
 \newcommand{\ee}{\end{eqnarray}}
\newcommand{\ov } {\over }

\begin{document}
\begin{titlepage}
\bigskip\begin{flushright}
UUITP-18/07\\
\end{flushright}
\begin{center}
\vskip 4cm
\Large{Entropy of String States at fixed Mass and Size}
\end{center}
\vskip 1cm
\begin{center}
\large{Diego Chialva \\ 
        Department of Theoretical Physics \\
        Uppsala University, Uppsala, Sweden \\
        \tt{diego.chialva@teorfys.uu.se}}

\end{center}
\date{}

\pagestyle{plain}
\vskip 3cm
\begin{abstract}
We provide
formulas for the entropy of free-string states depending on their mass,
charges and 
size, both in bosonic and superstring theory (IIA or IIB). We
properly define these quantities in 
full-fledged string 
theory. 
We then investigate the corrections to the entropy
due to self-interactions of the string for states with fixed mass, charge and
size, both for BPS and non-BPS configurations.
Again, the analysis is performed using string theory techniques.
\end{abstract}
\end{titlepage}
\newpage

\tableofcontents

\setcounter{section}{0}
\section{Introduction}

Black holes obey a set of laws that are formally identical to the
thermodynamical ones:
 \begin{enumerate}
 \item $\kappa_s= \text{constant}$ on horizon ($\kappa_s =$ surface gravity)
 \item $\delta M = {\kappa_s \ov 8 \pi G_N} \delta A+ \omega \delta J+\phi
  \delta q$

 ($M = \text{black hole mass},\, A=\text{horizon area},\, 
  \omega =\text{angular velocity at horizon},\,$

  $J= \text{angular momentum},\, \phi=\text{electric potential},\,
   q=\text{electric charge}$
  
 $G_N= \text{Newton's constant}$)
 \item $\delta A >0$\,\footnote{This law is violated by Hawking radiation,
 and therefore substituted by a generalized law stating that the
 area of the black holes plus the entropy of the universe do not
 decrease in a physical process.}
 \item $\kappa_s=0$ impossible for any physical process.
 \end{enumerate}

It is tempting to relate, then, $A$ to an ``entropy'' $S$, and $\kappa_s$ to a
temperature $T$. This is meaningless in classical theory, since black
holes do not radiate and it is not possible to associate a
temperature to them. At this level is therefore necessary a
quantum theory. In fact, Hawking radiation determines 
\[ \kappa_s = 2\pi T \] 
and then
\[ S = {A \ov 4 G_N}+ \ldots \]
where the dots stand for corrections due to higher
curvature terms and quantum contributions as well\footnote{The more general
classical formula for the entropy of a black hole, including higher
derivative terms, is Wald's one \cite{Wald}.}.

In the modern interpretation, we
think of classical gravitational black holes as a coarse-grained
description of a quantum system\footnote{This is one of the possible
  resolution of the problems posed by classical black holes. As an
  example it is typical of the "fuzzball" proposal, which sees the
  black hole solution as the coarse-grained description of many
  microstates' solutions which have no horizon. Other interpretations
  of black holes which do not includes processes of averaging and
  coarse-graining are in fact possible, but in general they have to
  cope with the problem of explaining how an horizon and trapped
  surfaces can arise from the pure quantum states accounting for the
  black hole microstates \cite{Amati}.}. This last description is the one that 
should provide us not only with an interpretation of black holes'
laws, but also with the possibility of deriving them from first
principles.

We will focus in particular on the entropy. According to Bekenstein
principle, the black hole's one is proportional to the area of the
horizon (plus corrections); for a quantum statistical ensemble, 
instead, the entropy
is defined as the logarithm of the number $G$ of microstates:
 \beq
 S = \ln(G).
 \eeq

Relating the two definitions represents the entropy issue. It is
necessary to individuate the correct ensemble of microstates accounting
for $S$: we need therefore a quantum gravity theory. 
String theory represents probably the best candidate nowadays for such
a theory and it has a general principle (known as the
``String-Black holes correspondence principle'') individuating those
microstates. In the case of supersymmetric BPS configurations at fixed
tree-level mass, their
counting at the level of free string (string coupling
$g_c = 0$) does not receive corrections, in the non
supersymmetric case, instead, quantum corrections must be taken in
consideration. 

In this work we want to focus on these effects on the
string theory side (microstates). We consider closed string, and perform our
analysis both in the bosonic and in the superstring theory (type IIB or IIA).
We will deal both
with states carrying no charges
and with states carrying charges of the Neveu-Schwarz type.
We begin in Section \ref{principle},
by reviewing the String-Black Holes Correspondence Principle. Special emphasis
is given to the role played by the value of the horizon radius of the
black hole and the corresponding requirements for the size of the
string microstates. 

It is therefore interesting to find the entropy of string microstates
depending on both mass and size (concept that we will define
properly in the following). This task is not easily solved because it
is not straightforward, in the quantum theory of strings, to define an
operator  
measuring the (average) size of string states. 
The main topic of this work will therefore be
to compute such entropy in a well-defined way.
Furthermore, the free-string entropy as a function of mass and size
will now receive corrections in both the non-BPS and the BPS cases
(renormalization of the radius).

We discuss and specify our statistical ensemble of closed string states
in section \ref{microensemble},
and in \ref{sizesec} we investigate the spatial distribution
and the number of microstates with zero charge at fixed squared mass and
``size'' . At the end of the
section we provide  formulas for the entropy of single free (bosonic
and super-)string states 
constrained both in mass and in size.

We extend our results to string states carrying
Neveu-Schwarz charges (winding and Kaluza-Klein mode numbers) 
in section \ref{stateswithcharges}, and study  BPS states as well.

In section \ref{massshiftsec} we study the
one-loop corrections: we propose a method for implementing the
constraint on the size of string states and investigate how this affects the
one-loop amplitude. The results are obtained by
evaluating full-fledged (super)string path integrals.
There are important differences between the non-BPS and the
BPS case: we elucidate them
and treat both cases, separately. Finally, we comment and conclude.

\section{The String-Black Holes correspondence principle}\label{principle}

String theory and black holes' physics set two
characteristic lenght scales:
 \begin{tabbing}
 ~~~~$R_{bh}$ \hspace{1.4cm} \=
 the black hole horizon radius 
 (Schwarzschild radius) \\
 ~~~~$l_s=\sqrt{\alpha'}$ \> the string lenght scale
 \end{tabbing}
so that
 \beq
 \begin{cases}
 \text{if}\quad R_{bh}\gg l_s & \text{general relativity description is
 reliable} \\
 \\
 \text{if} \quad R_{bh} \lesssim  l_s & \text{strings feel space-time as flat,
 $\alpha'$-corrections are important,}\\
                      & \text{string theory description is reliable}  \\
 \end{cases} \nonumber
 \eeq

The  ``String-Black Holes Correspondence
Principle'' states that a black hole is described by 
an ensemble of excited string
and/or D-brane states (depending on the type of charges the black hole
possesses) when $R_{bh} \sim  l_s$. 

There are two possible
interpretations of the Principle (for simplicity we consider now the
case without 
charges):
 \begin{itemize}
 \item a {\bf physical process} (Hawking radiation) where the black
 hole decreases its mass, therefore reducing the value of its
 Schwarzschild radius $R_{bh}\sim (G_N M)^{{1 \ov d-2}}$\footnotetext{We are in
$d=D-1$ spatial dimensions, and we relate Newton's constant
  to the string length
       as $G_N\sim g_c^2 
  (\alpha')^{d-1}$ at small closed string coupling $g_c$, see
   \cite{DamVenSelf}.} until $R_{bh}\sim l_s$ where a transition 
to an excited string states takes place;

 in this case
 $\begin{cases}
 g_c & \text{is fixed} \\
 M & \text{varies}
 \end{cases}
 $
 \item {\bf two complementary descriptions} valid in different regimes
 at equal mass;
 
 in this case
 $\begin{cases}
 g_c & \text{varies} \\
 M & \text{is fixed}
 \end{cases}
 $
 \end{itemize}

The possibility of equating the black hole entropy (proportional to a
power of its mass) and the string one (proportional to the square root 
of its mass)
relies on the fact that the first is constant in Plank units, the
second in string ones and therefore the entropies match at a
determined value of the string coupling. 
At the transition point it is found that, in units of $\alpha'$ 
\cite{HorPol96},
 \beq
 R_{bh} \sim l_s \Rightarrow g_c \sim M^{-{1 \ov 2}}
 \eeq
independently of the number of dimensions.
Since we are to consider very
massive string states, this value for the string coupling turns out to
be sufficiently small to allow perturbation theory.

We would expect that only states 
whose size is of 
the same order of the black hole horizon radius can be related to
the black hole at 
the transition point\footnote{Consider for example the gravitational binding to
mass ratio, or, for the case of the ``fuzzball" proposal, the fact that
the metric sourced by the microstates must differ from the one of a black hole
only at distance lower than the horizon radius.}. 
It is therefore interesting to find the entropy of string microstates
depending on both mass and ``size''.

In this work we
will determine the number (and therefore the entropy)
of perturbative (super)string states
depending on their mass and size at tree-level
and we will investigate the corrections to
their entropy due to the self-interactions of the string.
We will then consider states carrying charges, and extend the results
to that case.

In the past, a few attempts have been made to study
such issues:  \cite{DamVenSelf}, \cite{HorPolSelf} (see also
\cite{Cornalba}). In 
\cite{HorPolSelf}, it was
employed a thermal scalar formalism, interpreting the size of
the bound states of a certain scalar field as the size of the excited
string. The thermal scalar is a formal device capable to give us some
statistical information about the string system (string
gas). Nevertheless its relation to the string states remains
open. In 
particular, an Hamiltonian interpretation for its degrees of
freedom seems to question an identification between its states and the
string spectrum (see \cite{HagRidge}). 

The approach followed in \cite{DamVenSelf} was more directly
linked to a model of (bosonic open) strings, but the computations were
performed  
within a simplified toy model, believed 
to be valid in a large number of dimensions ($d \gg 1$), not taking
into account Virasoro constraints. 

We will perform our calculations in full-fledged string theory.

The computation of the entropy of strings
is ultimately connected also with the Hagedorn transition in
string theory, but we will deal with single-string
entropy and therefore our results do not apply directly.

Our conventions, here and in the following are:
 \be
 \alpha' = 4, && \quad D=d+1 \,\,\text{large space-time dimensions} \nonumber\\
  M_\text{tree}=\sqrt{N}&=&\text{bare mass of the string state} \nonumber\\
 g_c/g_o =&& \text{closed/open string coupling}. \nonumber
 \ee

Furthermore, objects with ``$c$'' subscript will refer to closed
strings, whereas those 
with an ``$o$'' subscript will relate to open strings.

\section{The microcanonical ensemble}\label{microensemble}

We want to determine statistical properties of massive string states,
in particular concerning their spatial distribution. We will use the 
{\em microcanonical} ensemble. Ensembles are defined by
density matrices: the microcanonical one has the
form\footnote{\label{operatorialdelta} The expressions for the density
  matrix are 
  meaningful when applied to the states of a system; with that understanding
  our notation with Dirac's delta functions is clear.}
 \beq 
 \rho_E = a_E\delta(E-\hat H)
 \eeq
where $\hat H$\,\footnote{From now on a $\hat ~$ will distinguish an
  operator from its value(s).} is the Hamiltonian of the system
and $a_E$ ensures the 
normalization of the density matrix:   
  \beq
  \text{tr}[\rho_E]=1
  \eeq
when traced over the states.

We can try to modify the traditional microcanonical ensemble,
fixing the value of other observables, in order to investigate different
statistical properties of the system. Considering a discrete observable 
with associated operator $\hat Q$, we can define the density
matrix:
 \beq
 \rho_{E, Q}= a_{E, Q} \delta(E-\hat H)\delta(Q-\hat Q).
 \eeq
The quantity 
 \be \label{numberstatesmicro}
  G(E, Q) & = & \text{tr}[\delta(E-\hat H)\delta(Q-\hat Q)] \nonumber \\
          & = & 
   \sum_\phi \langle\phi|\delta(E-\hat H)\delta(Q-\hat Q)|\phi\rangle
 \ee 
gives the number of states having the values $E, \, Q$ for the chosen
observables. It is, therefore:
 \beq
 a_{E, Q} = G(E, Q)^{-1}
 \eeq

If $Q$ represents a continuous observable, we need further to specify a
small interval $\delta Q$ (uncertainty) around the value of  the
observable we are
interested in, and define:
 \be
 \rho_{E, Q, \delta Q} & = & a_{E, Q, \delta Q} \, \delta(E-\hat H)\,
    \big(\theta(Q+\delta Q-\hat Q)-\theta(Q-\hat Q) \big) \nonumber \\
     & = & a_{E, Q, \delta Q} \, \delta(E-\hat H) \, 
       \int_{Q}^{Q+\delta Q}\delta(Q-\hat Q).
 \ee
Once again, tracing the density matrix over the states of the system,
yields the number $G(E, Q, \delta Q)=a_{E, Q, \delta Q}^{-1}$ of
microstates having values $E$ for the energy, 
and $Q < Q_i < Q+\delta Q$\footnote{Here, $i$ runs over the set of
  microstates.} for the other observable. 
We will let $\delta Q \to 0$, so that we can
write
 \beq
 \rho_{E, Q}  = G_{E, Q} \, \delta(E-\hat H)\delta(Q-\hat Q).
 \eeq
\newpage
Our microcanonical ensemble will be defined by fixing the values 
 \begin{itemize}
 \item of the level number operator $\hat N \equiv  -\hat p^2$   
 \item of the operator (to be defined)
    measuring the {\em size} of the string.
 \end{itemize}
In this way, we will be able to count the number of states $G(N, R)$
with fixed squared mass $N$ and size $R$, whose logarithm will
yield the entropy we are looking for.  

A problem arises:
it is difficult to define an
operator whose (average) value, when applied to a string state, 
represents its (average) size,
satisfying all the constraints of the
theory (superconformal or conformal constraints, or BRST constraints
in the various 
quantization procedures) or being computationally manageable (in
light-cone gauge), see \cite{DamVenSelf, SussSizeString}.
In order to cope with this problem, we follow a somehow
roundabout procedure. 

In sections \ref{sizesec}, \ref{stateswithcharges} we do this for free
string states, 
whereas in section \ref{massshiftsec}, we address
self-interacting strings.

\section{States with no charge} \label{sizesec} 

\subsection{The setup}\label{measuresetup}

As an illustration of the difficulties in defining an operator
to measure the {\em size} of a string state\footnote{We consider both pure
  and mixed states.}, let us discuss the most natural choice, which
will also clarify what we mean with the term {\em size}. Consider 
taking the quantized version of the classical average (squared) size
of a string: 
 \beq \label{classicalaverageradius}
 \hat R^2 = 
  {1 \ov \Delta\sigma_+\Delta\sigma_-}\int_0^{\Delta\sigma_+} \int_0^{\Delta\sigma_-}
    (X^O(\sigma_+, \sigma_-))^2 \quad \sigma_{\pm}=\sigma\pm\tau,
 \eeq
where $X^O$ represents the projection of the oscillator part of
the string coordinate
orthogonally to the center of
mass momentum of the string. Three evident issues regarding such operator are:
 \begin{itemize}
 \item its definition is gauge-dependent,
 \item the operator has a zero-order contribution
     proportional to 
     \beq
      \sum_{n=1}^\infty {1 \ov n}
     \eeq
  which needs to be interpreted and regularized (see \cite{SussSizeString}),
 \item the insertion of this operator in
       a path-integral is problematic because it does not commute with
       the BRST operator, or, when using Light-Cone gauge, poses ordering 
     problems\footnote{Consider the part of the operator proportional
       to $X^+ X^-$, where $X^-$ 
  is a quadratic function of the transverse coordinates.}. 
 \end{itemize}
  
The approach that we will choose solves all these issues. We will
adopt a physical way of measuring the spatial distribution of an
object: 
{\em the spatial distribution of an object
is obtained from the scattering of other (light) probes off it in an
elastic limit}. This will lead us to correctly define the density
matrix for our ensemble. 

\begin{figure}[htb]
 \begin{center}
\includegraphics[height=6cm]{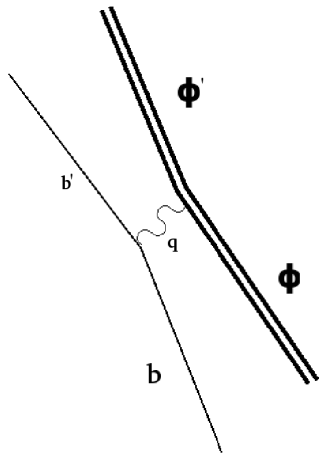}
\end{center}
\caption{{\small \sl{Scattering process $b+\phi \to b'+\phi'$. At low
      momentum transfer the process is dominated by the massless
      channel, which is represented here. }}} 
 \label{bornapprox}
\end{figure}

Consider, indeed, the Born approximation for the amplitude of a process
 \beq \label{scatproc}
 A=b+\phi \to b'+\phi', \quad q\equiv \text{exchanged momentum}
 \eeq
(see figure \ref{bornapprox}). At low momentum transfer, in the
 elastic limit, this is given by 
 \beq
 A \sim V(q^2) \cdots \sim {F_{b}(q^2) F_{\phi}(q^2) \ov q^2}
 \eeq
and $F_i(q^2), \, m_i$ can be interpreted respectively as the {\em form
 factor} and the mass
of the particle $i$.
In particular, for elastic scattering (when $q^2= \vec q^{~2}$),
the form factor represents the Fourier transform of the spatial
 distribution $\mu_i(x)$:
 \beq
 F_i(\vec q^{~2})= \int d^d x e^{-i \vec q\cdot \vec x}\mu_i(\vec x)
 \eeq

According to the nature of the probe $b$, we obtain different distributions: 
mass distribution, charge distribution \ldots.
In the following we will consider the {\em mass distribution},
choosing $b$ to be a graviton (actually a superposition of
graviton, dilaton and Kalb-Ramond field). 
Our targets $\phi$ will be closed string states extended in the large
uncompactified dimensions, though the
analysis can be 
carried over for open strings as well\footnote{See
  \cite{MitchSundSize} for a study of the size of open 
  string states on the leading Regge trajectory using tachyons and
  photons as probes.}.

\subsection{The String Spatial Distribution.}\label{SpatialDistribution}

The string theory formula\footnote{Note
  that this formula, and therefore the result that we will obtain, as
  we will discuss in the following are very different from the
  formulas in \cite{ManesFF1}. In our case the interpretation of $F_N(q^2)$ as
  a form factor is justified, according to scattering theory, 
  whereas in \cite{ManesFF1} it
  could not be accepted, and indeed a different
  interpretation of the results was proposed in \cite{ManesFF2}}
for the amplitude (S-matrix) for the process (\ref{scatproc}) is:
 \beq \label{borampli}
 A_{\text{closed}}=g_c^2\int d^2\,z \, \langle \phi|V(k', 1) V(k, z)|\phi\rangle
 \eeq

Let us discuss for the moment an ensemble of states with
fixed squared mass $N$ only. We then need to trace over the density
matrix\footnote{$G_c(N)$ is the number of states at mass level $N$,
  the trace is over {\em physical} string states.} 
 \beq \label{massproject}
 \rho_N = {1 \ov G_c(N)}\, \delta(N -\hat N)=
  {1 \ov G_c(N)}\oint {dw \ov w^{N+1}}w^{\hat N}
 \eeq
so that:
 \beq
 A_{\text{closed}}=g_c^2\int d^2\,z \, tr[V(k', 1) V(k, z)\rho_N \tilde \rho_N] ,
 \eeq
where objects with a tilde refer to the anti-holomorfic (left-moving) sector.

We consider separately the cases of the bosonic string and the superstring.
We will find out that the computations are very similar. We
therefore discuss  
first and at length the superstring, and leave the bosonic one at the end.

\paragraph{The superstring.}
~~

As we said, our probe consists of a superposition of the graviton, the dilaton
and the Kalb-Ramond field, represented by the vertex operator:
 \beq \label{gravvertex}
 V(k, z)= {2 \ov \alpha'} e^{i k\cdot X} 
          (\xi\cdot\partial X-{i \ov 2} \xi\cdot \psi k\cdot\psi)
          (\tilde\xi\cdot\bar\partial X-{i \ov 2}
              \tilde\xi\cdot \tilde\psi k\cdot\tilde\psi)
 \eeq
where we have (formally) written the polarization tensor as
$\xi_{\mu\nu}=\xi_\mu \tilde\xi_\nu$ and $X, \psi$ are, respectively, the
space-time string bosonic and fermionic coordinates. We perform our
computations in the gauge $\xi_{00} = 0$.  

We will make use of the relation (Kawai, Lewellen and Tye,
\cite{KLT})\footnote{Here we have explicitly written the $\alpha'$
  squared lenght in order to present clearly the formula. Remember
  that eventually in the computations we will always set $\alpha'=4$.}:
 \beq
 A_c(1234; \alpha', g_c)={\pi i g_c^2 \alpha' \ov g_o^4}
 \sin(\pi\alpha't) A_o(s, t; {\alpha' \ov 4}, g_o) \tilde A_o(t, u; {\alpha' \ov 4}, g_o).
 \eeq
where  $s, t, u$ are Mandelstam variables. 

The amplitude that we will compute is therefore:
 \beq \label{borampliop}
 A_o(s, t; 1, g_o)=  g_o^2\int dy
   \,\text{tr}[V_{\text{open}}(k', 1)V_{\text{open}}(k, y)\rho_N]
 \eeq
with
 \beq
 V_{\text{open}}(k, y) = {1 \ov \sqrt{ 2\alpha'}} e^{i k\cdot X(y)} 
          \left(i y \xi\cdot\partial_y X(y)+2\alpha'  k\cdot\psi(y) \xi\cdot \psi(y)\right).
 \eeq

We will consider the limit $t\equiv -q^2=-(k+k')^2 \to 0$. 
The lowest terms of the amplitude in this limit can be calculated using 
the OPE (see \cite{ManesFF1}):
 \beq \label{OPE}
 V_{\text{open}}(k', 1) V_{\text{open}}(k, y) \underset{y\to 1}{\sim} 2\xi\cdot\xi' \text{{\small $(1-q^2)\,
     (1-y)$}}^{\text{{\tiny $2 k'\!\!\cdot \!\!k\!\!-\!\!2$}}}y^{\text{{\tiny $2 k\!\!\cdot \!\!\hat p$}}}
      e^{i q\cdot \hat X_O(1)} e^{iq\cdot \hat x}
 \eeq
where $\hat X_O$ indicates the oscillator part of $X$. The
amplitude factorizes as:
 \beq
 A_{\text{open}} =g_o^2 A_o^{\text{zero modes}}A_o^{\text{oscillators}}
 \eeq
By writing $y = e^{-\epsilon}$ with $\epsilon\to 0 $, we find the 
result\footnote{We need to perform the same analytical continuation as
 for the Veneziano 
amplitude, as usual in these representation of the string amplitudes.}:
 \be \label{integralGammaSuper}
 A_o^{\text{zero modes}} & = &
     -\int d\epsilon \,\epsilon^{q^2-2}e^{-\epsilon(2 k\cdot p+1)}\, (1-q^2)
      \\
   & & \underset{q^2 \to 0}{\sim} {(2\sqrt{N}E)^2 \ov q^2} \sqrt{F_b(q^2, E)} \, (2\sqrt{N})^{-q^2}.
 \ee
where we have defined $F_b(q^2, E) \equiv e^{-2q^2\ln(E)}$ and $E\equiv k^0$.

Therefore:
 \beq
 A_c(1234; 4, g_c)\sim \pi^2 i g_c^2 
  {(2\sqrt{N}E)^4 \ov q^2} F_b(q^2, E) (2\sqrt{N})^{-2q^2} 
  A_o^{\text{oscillators}} \tilde A_o^{\text{oscillators}}
 \eeq
with
 \beq \label{Aopenoscill}
  A_o^{\text{oscillators}} = \text{tr}[e^{i q\cdot X_O(1)} \rho_{N}]
 \eeq
and we have expanded $\sin(-\pi t)\sim -\pi t \sim \pi q^2$. 

According to the results and the discussion in \cite{ ManesFF1,
  SussBlStrCompl}, we identify  
$F_b(q^2, E)$ with the form factor for the probe $b$.

It is now straightforward to read the form factor for the target
$\phi$ at squared mass $N$:
 \be \label{formfactorN}
 F_N(q^2) & = & N^{-q^2} A_o^{\text{oscillators}} \tilde A_o^{\text{oscillators}} 
 \\ 
 & = &  {N^{-q^2} \ov G_c(N)} 
     \oint {dw \ov w^{N+1}} \oint {d\tilde w \ov \tilde w^{N+1}}
   {g(w)g(\tilde w) \ov (f(w)f(\tilde w))^{d-1}}
 e^{-2q^2\sum_{n=1}^\infty{w^n \ov n(1-w^n)}+{\tilde w^n \ov n(1-\tilde w^n)}}
 \nonumber 
 \ee
where 
 \beq \label{fermiondet}
  g(w)= \left({1\ov
                       \sqrt{w}}g_3(w)^{d-1}-{1\ov
                       \sqrt{z}}g_4(w)^{d-1}+
                       g_2(w)^{d-1}\right)\, \ \ \ 
 \eeq
 \be
 f(w)= \prod_{n=1}^\infty (1-w^n) &&
 g_3(w)=\prod_{r={1\ov 2}}^\infty(1+w^r) \\
 g_4(w)=\prod_{r={1\ov 2}}^\infty(1-w^r) &&
 g_2(w)=\prod_{r=0}^\infty(1+w^r) \ .
 \ee
We compute the loop-integrals by saddle point approximation for large $N$,
finding
 \beq \label{saddlepointsize} 
 \ln(w) \sim - {\pi \ov \sqrt{N}}\sqrt{{d-1 \ov 4}-{q^2 \ov 3}}
 \eeq
and similarly for $\tilde w$.

Therefore, considering the elastic limit:
 \be
 F_N(\vec q^{~2}) & \underset{N\to\infty}{\sim} & 
  {e^{4\pi\sqrt{N}\sqrt{{d-1 \ov 4}-{\vec q^{~2} \ov 3}}} \ov G_c(N)}\pi^{d}
   \left({(d-1) \ov 4 }-{\vec q^{~2} \ov 3}\right)^{{d \ov 2}} N^{-{d+2 \ov 2}}  
  \nonumber \\
   & \underset{\substack{\vec q^{~2}\to 0 \\ \\ N \to \infty}}{\sim} & 
   e^{-{4\pi \ov 3} \sqrt{{N \ov D-2}} \vec q^{~2}}
 \ee
where in the last line we have simplified the result with 
 \beq
   G_c(N) \sim e^{2\pi\sqrt{N\,(d-1)}}\pi^{d}
      \left({d-1 \ov 4}\right)^{{d \ov 2}} N^{-{d+2 \ov 2}}.
 \eeq
Finally, the average radius is
 \beq
 \langle r^2 \rangle = -2\, d \,
   \partial_{\vec q^{~2}} F_N(\vec q^{~2})|_{\vec q^{~2}=0} 
 =  {8\pi \,d  \ov 3} \sqrt{{N \ov d-1}}
 \eeq
and the mass distribution
 \beq \label{sizedistribution}
 \mu_N(\vec x) = 
  {1 \ov (2\pi)^d} \int d^dq e^{i \vec q \cdot \vec x} F_N(\vec q^{~2})=
 \left({3 \ov 16\pi^2} \sqrt{{d-1\ov N}} \right)^{{d \ov 2}}
 e^{-{3 \ov 16 \pi}\sqrt{{d-1\ov N}} \vec x^2}.
 \eeq

\vskip 0.4cm

\paragraph{The bosonic string.}
~~
\nopagebreak

The case of the bosonic string follows the same steps. 
A few things are different:
 \begin{itemize}
 \item the vertex operator for the probe now is:
    \beq \label{gravvertexbosonic}
      V_{\text{open}}(k, y) = {1 \ov \sqrt{ 2\alpha'}} e^{i k\cdot X(y)} 
          \left(i y \xi\cdot\partial_y X(y)\right).
    \eeq
 \item due to the absence of fermionic excitations, the 
   number of closed string states at fixed large mass squared $N$ is
   \beq
     G_c(N) \sim e^{4\pi\sqrt{N}\sqrt{{d-1 \ov 6}}}\pi^{d}
    \left({(d-1) \ov 6 }\right)^{{d \ov 2}} N^{-{d+2 \ov 2}};
   \eeq
  \item the integral (\ref{integralGammaSuper}) becomes now:
   \beq \label{integralGammaBosonic}
    A_o^{\text{zero modes}}  = 
     -\int d\epsilon \,\epsilon^{q^2-2}e^{-\epsilon(2 k\cdot p+1)}\,.
   \eeq
  \end{itemize}
Namely, we see that the integral would be divergent also for $q^2=1$, 
corresponding to the exchange of a tachyon. But we are considering the limit 
$q^2 \to 0$, picking out the graviton pole, so that
 \beq 
 A_o^{\text{zero modes}}  
 \underset{q^2 \to 0}{\sim} {(2\sqrt{N}E)^2 \ov q^2} \sqrt{F_b(q^2, E)} \, (2\sqrt{N})^{-q^2}
 \eeq
as for the superstring. 
Therefore we obtain:
 \begin{itemize}
  \item form factor
     \beq
       F_N(\vec q^{~2}) \underset{{\vec q^{\,2}\ov N}\to 0}{\sim}  
         e^{-2\pi \sqrt{{2 \,N \ov 3\, (d-1)}} \vec q^{~2}},
     \eeq
   \item the average radius 
     \beq
      \langle r^2 \rangle =   
       4\sqrt{2}\,\pi \,d  \sqrt{{N \ov 3\,(d-1)}},
     \eeq
   \item the mass distribution
     \beq \label{sizedistributionbos}
      \mu_N(\vec x) = 
       \left({1 \ov 8 \pi^2}\sqrt{{3\,(d-1) \ov 2N}} \right)^{{d \ov 2}}
       e^{-{1 \ov 8 \pi}\sqrt{{3\,(d-1) \ov 2N}} \vec x^2},
     \eeq
 \end{itemize}
being intended that
the number of extended spatial dimensions now can go up to $d=25$, not
only up to 9 as for the superstring.

\subsubsection{Corrections}

We show here how the lowest terms in the expansion  
for $y \to 1$ in (\ref{OPE}), dominate the amplitude (\ref{borampliop})
and the result is safe against possible corrections 
in a determined kinetic and mass range. 

\paragraph{The superstring.}
~~

Without any approximations, the amplitude (\ref{borampliop}) is given
by
 \be
 A_o(s, q^2;1, g_o) & \sim & {g_o^2 \ov G_o(N)}\!\int\!\!d\epsilon 
  \!\oint\!\!{dw \ov w^{N+1}}
  {g(w) \ov f(w)^{d-1}}\,\,
  e^{-\epsilon (2k\cdot p+1)}\,\,\psi(\epsilon, w)^{q^2}\,  \\
  & & ~~~~~~~~~~~~~~~~~~  \times 
 \left[-2\partial^2\epsilon \ln{(\psi(\epsilon, w))} + \chi(\epsilon, w)\right]
 \nonumber
 \ee
with
 \be \label{logterm}
 \psi(\epsilon, w) & = & 
  (1-e^{-\epsilon})
 \prod_{n=1}^\infty e^{-q^2 {w^n \ov n(1-w^n)} (e^{n\epsilon}+e^{-n\epsilon})} \\
 \partial^2_\epsilon\ln{(\psi(\epsilon, w))} & = & 
  \sum_{n=1}^\infty n e^{-\epsilon n} +
 \sum_{n=1}^\infty {n w^n \ov (1-w^n)}(e^{n\epsilon}+e^{-n\epsilon})  \\
 \chi(\epsilon, w) & = & 2q^2 
 \sum_{s=2}^4 \left( {\theta_s(0) \ov \theta_2(0)}\right)^{{d-1 \ov 2}} 
 {\theta_s(\epsilon)^2 \theta_1'(0)^2 \ov\theta_1(\epsilon)^2\theta_s(0)^2} ,
 \ee
where we have written $y=e^{-\epsilon}$. Note that $\theta_s(z) \equiv
\theta_s({z \ov 2\pi i}, {\ln(w) \ov 2\pi i})$ in the usual notation,
where the $\theta_s$'s are the 
Theta functions.

Expand for $\epsilon \to 0$: 
 \be
 I_\epsilon & \sim & \int d\epsilon e^{-\epsilon\,(2k\cdot p+1)} \epsilon^{q^2-2}
  e^{-2q^2\sum_n {w^n \ov n (1-w^n)}}
  \left(1-q^2+O\left({w\epsilon^2 \ov (1-w)}\right)\right)
  \nonumber \\
  & \sim & (2k\cdot p)^{-q^2+2}\Gamma(q^2)\left(1+O({1\ov (k\cdot p)^2})\right). 
 \ee
 Being $k\cdot p =-E\sqrt{N}$ ($E$ probe energy,
 $\sqrt{N}$ tree-level mass for the massive 
 state), our results
 appear to be correct in the limit of heavy massive string states and
 probes at high energy.

\vskip 0.4cm
\paragraph{The bosonic string.}
~~

The bosonic string case is similar to the superstring one: it can be
obtained eliminating  
from the formulas above the term $\chi(\epsilon, w)$ and substituting 1
to $g(w)$, which leads to the
result:
 \beq
 I_\epsilon \sim  (2k\cdot p)^{-q^2+2}\Gamma(q^2-1)\left(1+O({1\ov (k\cdot p)^2}) \right). 
 \eeq
showing again the validity of our expansion for heavy target states
and probes at high energy.

\subsection{Number of string states of a given mass and size.}
\label{numberstatessection}

Ultimately, we are
interested in (the logarithm of) the number of states with a given mass
and size. This 
can be obtained from the results in section
\ref{SpatialDistribution}. Indeed, looking at formulas  
(\ref{massproject}, \ref{Aopenoscill}, \ref{formfactorN},
\ref{sizedistribution}) and remembering that we are considering
elastic scattering, we can write: 
 \beq \label{formfactorexpform}
 F_N(\vec q^{~2}) = {1 \ov G_c(N)}
  \text{tr}[e^{i\vec q\cdot \hat{\vec{ X}}^O_{\text{closed}}(1)} \delta(N-\hat N)]
 \eeq
where
 $
 \hat{\vec{ X}}^O_{\text{closed}}(z)
 $
is the projection of the oscillator part of the string coordinates
operator, orthogonally to the momentum of the string\footnote{In the
  limit $\vec q^2 \to 0$, the state $|\phi\rangle$ is in his rest frame.}.

Therefore
 \beq \label{massdistribdeltaform}
 \mu_N(\vec x)= \int {dq \ov (2\pi)^d} e^{-i \vec q\cdot \vec x}F_N(q^2)= {1 \ov G_c(N)}
  \text{tr}[\delta(\vec x-\hat{\vec{X}}^O_{\text{closed}}(1)) \delta(N-\hat N)]
 \eeq

For fixed $\vec x$, we recognize in $\mu_N(\vec x)$ the trace of the
(incorrectly normalized) density matrix for an ensemble with fixed
$\vec X^O, \hat N$. 
We note that, in terms of the string oscillators $\vec \alpha_m,
\,\vec{\tilde{\alpha}}_m $: 
 \beq
 \hat{\vec{X}}^O_{\text{closed}}(1) =
 \sqrt{2}\,\sum_{\substack{m=-\infty\\m\neq 0}}^\infty {\vec \alpha_m \ov m}+
 {\vec{\tilde{\alpha}}_m \ov m}
 \eeq
and, thanks to the normal ordering in the string amplitude, $\vec q^{~2}$ is
multiplied by 
\beq
 2\sum_{m=1}^\infty{\vec \alpha_{-m}\cdot\vec \alpha_m \ov m}+
 {\vec{\tilde{\alpha}}_{-m}\cdot\vec{\tilde{\alpha}}_m \ov m}= :\hat R^2: \,,
 \eeq
where $\hat R^2$ is the operator, modulo the zero-point
contribution\footnote{The zero-point contribution gives rise to the
  factor $N^{-q^2}$ in (\ref{formfactorN})
which is negligible for $N \to \infty$.},
which we had described in (\ref{classicalaverageradius}), and $::$
denotes normal ordering.

Therefore,  
writing $\vec x$ in spherical coordinates and integrating over the
angular dependence,  
we obtain:
\begin{itemize}
 \item for the {\bf superstring}
 \begin{itemize}
  \item the number of closed string states with fixed $R, N$
   \beq \label{numberfixedNR}
    G_c(N, R)\!\! =  \!\!{2 \ov \Gamma({d \ov 2})}
     \left({3 \sqrt{d-1} \ov 16\pi^2 \sqrt{N}} \right)^{\!{d \ov 2}} \left({R \ov \sqrt{N}}\right)^{\!d\!-\!1} 
   {e^{\pi\sqrt{d-1}\left(2\sqrt{N}-{3 \ov 16 \pi^2\sqrt{N}} R^2\right)}\ov N^{3 \ov 2}}
   \eeq
  \item and the entropy
   \be
     S & = & \ln(G_c(\sqrt{N}, R)) \nonumber \\
      & \sim & 2\pi \sqrt{N}\sqrt{d-1}-{3\sqrt{d-1} \ov 16\sqrt{N} \pi}  R^2 + 
       \ln\left({R^{d-1} \ov \sqrt{N}^{{3\ov 4}d+1}} \right)
   \ee
  \end{itemize}
 \item for the {\bf bosonic string}
 \begin{itemize}
  \item the number of closed string states with fixed $R, N$
   \beq \label{numberfixedNRbos}
    G_c(N, R)\!\! = \!\!{2 \ov \Gamma({d \ov 2})}
     \left({\sqrt{3 (d-1)} \ov 8\sqrt{2}\pi^2 \sqrt{N}} \right)^{\!\!{d \ov 2}}\!\!\!\! \left({R \ov \sqrt{N}}\right)^{\!\!d\!-\!1} 
  \!\! {e^{\pi\sqrt{d-1}\left(\sqrt{{8N \ov 3}}-{\sqrt{3} \ov 8 \sqrt{2}\, \pi^2\, \sqrt{N}}  R^2\right)}\ov N^{3 \ov 2}}
   \eeq
  \item and the entropy
   \be
     S & = & \ln(G_c(N, R)) \nonumber \\
      & \sim & 4\pi \sqrt{N}\sqrt{{d-1\ov 6}}-
      {\sqrt{3\,(d-1)} \ov 8\sqrt{2}\,\sqrt{N}\, \pi}  R^2 + \ln\left({R^{d-1} \ov \sqrt{N}^{{3\ov 4}d+1}} \right).
   \ee
  \end{itemize}
\end{itemize}

\vskip 0.4cm

Two remarks are important at this point. First, looking at
(\ref{formfactorexpform}, \ref{massdistribdeltaform}), we note that we
have been inserting an operator
 \beq
 \delta(\vec x-\hat{\vec{ X}}^O_{\text{closed}}(1))= 
  \int d\vec q \, e^{i \vec q\cdot \vec x-iq\cdot\hat{\vec{ X}}^O_{\text{closed}}(1)}
 \eeq
in a string path integral. This operator, being integrated over all
momenta, is off-shell. But string theory is defined only on-shell, how
is then possible that our computation is correct? We appreciate here,
the importance of the factorization property of (string) amplitudes:
factorizing two external legs of an amplitude, the momentum square $q^2$ flowing
along the connecting propagator is a variable, allowing 
analytic continuation.

Since $[\hat{\vec{X}}^O_{\text{closed}}(1), \hat N] \neq 0$, we could
also wonder whether our computation 
for the number of states is incorrect, because the result in
(\ref{numberfixedNR}, \ref{numberfixedNRbos}) 
should be independent from the ordering of the two deltas. Naturally,  
$\delta(\vec x-\hat{\vec{ X}}^O_{\text{closed}}(1)) \delta(N-\hat
N)$ and $\delta(N-\hat N)\delta(\vec x-\hat{\vec{X}}^O_{\text{closed}}(1))$ 
yield the same result when traced over, and,
furthermore, we are working with very massive
string states,  
for which it is also reasonable to take a semi-classical limit.

\section{States carrying Neveu-Schwarz charges} \label{stateswithcharges}

The results obtained in the previous sections can be extended to
ensembles of string states carrying Neveu-Schwarz charges $Q_R, Q_L$. 
We have to distinguish states according to their mass and their
winding and Kaluza-Klein mode numbers $(m^i, n^i)$, such that:
 \be
 Q^i_{R, L} & = & \left({n^i \ov r^i}\pm {m^i r^i \ov 4}\right) \\
 Q^2_{R, L} & = & \sum_i Q_{R, L}^{i\,2},
 \ee
where $r^i$ is the radius\footnote{Recall that we set $\alpha'=4$
  and express everything in units of $\alpha'$.} of
 compactification in the $i$-th
 compactified direction.

The mass-shell condition and the Virasoro constraint $L_0-\tilde
L_0=0$ read:
 \be
 M^4 & = & Q_L^2 +N_L \\
     & = & Q_R^2 +N_R \\
 N_L-N_R & = &\sum_i n^im^i.
 \ee
where $L, R$ indicate respectively the holomorphic and anti-holomorfic sectors.

We define our microcanonical system by fixing charge and squared mass,
or, more conveniently and equivalently, by constraining the values of the
operators:
 \beq
 \hat N_L= -\hat p^2-\hat Q_L^2 \quad  \hat N_R= -\hat p^2-\hat Q_R^2.
 \eeq
and letting their values, $N_L, N_R$, be large.
Therefore:
\begin{itemize}
 \item for the {\bf superstring}
 \begin{itemize}
  \item the number of closed string states with fixed size, mass,
   charge is
   \be \label{numberfixedNRcharge}
    G_c    \sim {2\ov \Gamma({d \ov 2})}
     \left({3 \sqrt{d-1} \ov 8\pi^2 \mathcal{N}} \right)^{{d \ov 2}}
    \left({R \ov N_L^{{1 \ov 4}}N_R^{{1 \ov 4}}}\right)^{d-1} 
     {e^{\pi\sqrt{d-1}\left(\mathcal{N}-{3 \ov 8\pi^2\mathcal{N}}R^2\right)}\ov N_L^{{3 \ov 4}}N_R^{{3 \ov 4}}}
   \nonumber \\ 
   \ee
  \item and the entropy
   \be
     S & = & \ln(G_c) \nonumber \\
      & \sim & \pi \mathcal{N}\sqrt{d-1}-{3\sqrt{d-1} \ov 8\mathcal{N} \pi}R^2 
   +\ln\left({R^{d-1} \ov N_L^{{d+2 \ov 4}}N_R^{{d+2\ov 4}}\mathcal{N}^{{d \ov 2}}} \right)
   \ee
  \end{itemize}
 \item for the {\bf bosonic string}
 \begin{itemize}
  \item the number of closed string states with fixed size, mass,
   charge is
   \beq \label{numberfixedNRboscharge}
    G_c = {2 \ov \Gamma({d \ov 2})}
     \left({\sqrt{3 (d-1)} \ov 4\sqrt{2}\pi^2 \mathcal{N}} \right)^{{d \ov  2}} 
    \left({R \ov N_L^{{1 \ov 4}}N_R^{{1 \ov 4}}}\right)^{d-1} 
     {e^{\pi\sqrt{d-1}\left(\sqrt{{2 \ov 3}}\mathcal{N}-{\sqrt{3} \ov 4 \sqrt{2}\,\pi^2\, \mathcal{N}}  R^2\right)}\ov N_L^{{3 \ov 4}}N_R^{{3 \ov 4}}}
   \eeq
  \item and the entropy
   \be
     S & = & \ln(G_c) \nonumber \\
      & \sim & 2\pi \mathcal{N}\sqrt{{d-1\ov 6}}-
      {\sqrt{3\,(d-1)} \ov 4\sqrt{2}\,\mathcal{N}\, \pi}  R^2 + \!\!
   \ln\!\!\left({R^{d-1} \ov N_L^{{d+2 \ov 4}}N_R^{{d+2 \ov 4}}\mathcal{N}^{{d \ov 2}}} \right).
   \ee
  \end{itemize}
\end{itemize}
where we have written:
 \beq
 \mathcal{N}=\sqrt{N_L}+\sqrt{N_R}.
 \eeq

\subsection{BPS states} \label{BPSentropyfree}

We study, now, BPS configurations of fundamental superstrings.
They are states with:
 \beq
 M^2=Q^2_L, \quad N_L=0, \quad N_R=\sum_i n^im^i.
 \eeq
We find:
 \begin{itemize}
  \item the number of BPS string states with fixed size, mass,
   charge is
   \be \label{numberfixedNRBPS}
    G_c    \sim {2\ov \Gamma({d \ov 2})}
     \left({3 \sqrt{d-1} \ov 8\pi^2 \sqrt{N_R}} \right)^{{d \ov 2}}
    \left({R \ov N_R^{{1 \ov 4}}}\right)^{d-1} 
     {e^{\pi\sqrt{d-1}\left(\sqrt{N_R}-{3 \ov 8\pi^2\sqrt{N_R}}R^2\right)}\ov N_R^{{3 \ov 4}}}
.  \nonumber \\ 
   \ee
  \item and the entropy
   \be
     S & = & \ln(G_c) \nonumber \\
      & \sim & \pi \sqrt{N_R}\sqrt{d-1}-{3\sqrt{d-1} \ov 8\sqrt{N_R} \pi}R^2 
   +\ln\left({R^{d-1} \ov N_R^{{d+1 \ov 2}}} \right).
   \ee
  \end{itemize}

It is interesting to note that the average radius for this ensemble is
 \beq \label{sizeBPS}
 \langle r^2 \rangle = {4\pi \,d  \ov 3} \sqrt{{N_R \ov d-1}}.
 \eeq

\section{One-loop corrections} \label{massshiftsec}

\subsection{States with no charge}

The counting of states at a given mass level is affected by
the self-interaction of the string, unless we are considering  
supersymmetric configurations, which enjoy a protection
mechanism for the mass. We are interested in counting at fixed mass
and size, and therefore also the supersymmetric case will receive
corrections.

Studying the mass-shift of fundamental closed strings means
to compute one-loop amplitudes with the insertions of two vertex
operators representing the string state. Such calculations are
difficult to be performed and even defined in string theory for a
series of reasons:
 \begin{itemize}
 \item the form of vertex operators for massive states is complicated
 \item looking for statistical properties means in principle to be
   able to compute one-loop two-points amplitudes for all possible string states
   in an ensemble,  
   but only a few vertex operators are explicitly known
 \item one-loop two-points amplitudes are divergent (due to the presence
     of an imaginary part); they need analytic continuation, but
     String Theory is defined only on-shell. 
 \end{itemize}
 
An optimal method for solving these problems and computing would be
factorization (\cite{SundbShift}): starting from a known four-point
amplitude, we can factorize the external legs pairwise and obtain the
mass-shifts for the intermediate states as the residue of the double
pole for the center of mass energy. In that case we do not need the
detailed knowledge of the form of vertex operators and, as we said
above, the momentum square
flowing in the loop is now a variable, allowing analytical
continuation. Unfortunately this approach has a residual
problem: in order to identify mass-shifts for the
various states we need to know the form of all their couplings with
the external legs of the amplitude (for
particular states, namely those on the Regge trajectory, which are 
non-degenerate, the method
works, see \cite{SundbShift}). 

We are interested in the mass renormalization for
states with both mass and size fixed. The idea is that the formulas
for the entropy
obtained in sections \ref{sizesec} and \ref{stateswithcharges}
will receive corrections, such that
string states would have a typical
size\footnote{Given by the saddle points of the integral
  \beq 
 G_c(M) = \int d R  e^{S(M, R)},
 \eeq 
where  $G_c(M)$ is the total number of states at fixed squared mass
$M^2$.} matching the radius of the correspondent black hole at the
transition point.

The average mass-shift for states constrained in both mass and
(average squared) size
can be written as:
 \beq \label{masshisftfixedNR}
 \Delta M_{N, R} = {M^{-1} \ov G_c(N, R)} 
  \text{tr}[\widehat{\Delta M}^2\, \delta(N-\hat N)\delta(R^2-\hat R^2)]
 \eeq
where $\widehat{\Delta M}^2$ is an operator yielding the squared mass
shift once 
applied to a  
set of states\footnote{It can be obtained opportunely normalizing the
real part of the one-loop S-matrix operator.}. 

Once again there is an issue in defining an operator for the
observable ``size''; we try to cope with this
by
relying on our factorization
procedure, as in sections \ref{sizesec}, \ref{stateswithcharges}.
We consider therefore the one-loop amplitude for two states
represented by vertex operators $V_\phi$, in the appropriate pictures
for the superstring case,
and two probes with 
vertex operator $V$ given in (\ref{gravvertex}) or the analog for the
bosonic string case. 
Our goal is to factorize the full amplitude 
so that it can account for a mass-shift amplitude with the insertion
of a delta function  
constraining the operator $\hat{\vec{ X}}^O_{\text{closed}}$ 
as in (\ref{massdistribdeltaform}). We argue that, for the same
reasons explained in section \ref{numberstatessection}, this will be
the most efficient way to constrain the size of the ensemble states.
The relevant result, in the limit of 
low momentum transfer $q \equiv k+k'$
and elastic scattering, 
can be obtained from the OPE and integration over $\xi, \bar\xi$, as
follows\footnote{To avoid cluttering formulas we will write
  $V_\eta(v)$ instead than $V_\eta(v, \bar
  v)=\mathcal{V}_{\eta}(v)\mathcal{V}_{\eta}(\bar v)$ for the vertex
  operator corresposnding to a state $|\eta\rangle$. Furthermore, we
  have avoided, in formula (\ref{massshiftfixingNR}) to write the sign
  of integration over $\xi, \bar\xi$ on the left hand side.}:
 \beq \label{massshiftfixingNR}
 \langle V_{\phi'}(p', 0) V(k', z) V(k, \xi) V_\phi(p, \nu)\rangle_{T^2}\!\!
 \underset{\substack{q^2 \to 0 \\ \xi \sim z}}{\sim}\!\!{1 \ov \vec q^{\,2}}\,
 \langle V_{\phi'}(p', 0)\,e^{i\vec q\cdot \hat{\vec{ X}}^O_{\text{closed}}(z)}|z|^{\!\!4k'\cdot \hat p}V_\phi(p, \nu)\rangle_{T^2}
 \eeq

\begin{figure}[htb]
 \begin{center}
\includegraphics[height=6cm]{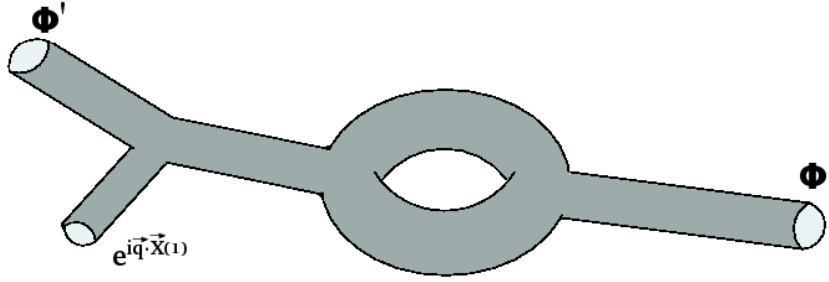}
\end{center}
\caption{{\small \sl{The diagram for the string amplitude for the one-loop mass
      renormalization at fixed $\hat{\vec{ X}}^O_{\text{closed}}(1)$. }}} 
 \label{massshiftfixedNRfigure}
\end{figure}

We want to discuss mass renormalization, but the amplitude involving
(\ref{massshiftfixingNR}) 
corresponds to various field theory
diagrams, accounting also for vertex corrections. To obtain
the one relevant for the mass-shift, we propose to consider the limit $z
\to \nu$, 
in order to single out the string amplitude represented in
figure \ref{massshiftfixedNRfigure}. 
We must be careful to take a limit where
the vertex operators $V_\phi(1), V(z), V(\xi)$ approach. Indeed when they
do it altogether
at the same rate, we are in a situation that leads to a dangerous
infrared divergence 
(see \cite{Weinberg:1985je, Seiberg:1986ea, Sen:1987bd}).
Instead, we look for a limit where two of them
($V(z), V(\xi)$) 
first approach each other, and then, at a slower rate,
$V_\phi(\nu)$.

It is not necessary for our purposes, but it simplifies the formulas,
if
we work in a {\em time gauge} (\cite{Hwang,
  Evans, MaldaLiuLamb, JMKeski}), such that:
 \begin{align} \label{genericvertex}
 V_\phi & =  e^{-i p^0 X^0+i \vec p\cdot \vec X}
         V(X^i) & \text{for the bosonic string}, \\
 V_\phi & =  e^{-i p^0 X^0+i \vec p\cdot \vec X}
         V(X^i, \psi^i, S^\alpha) & \text{for the fermionic string}. 
 \end{align}
Here the $S^{\alpha}$ are spin-fields, $i$ runs over the spatial
dimensions and $-p^2=N$. 
Then:  
 \be \label{oneloopfixedRN}
 A^R_{T^2} & = & g_c^4
  \iiint\langle V_{\phi'}(0) e^{i\vec q\cdot \hat{\vec{X}}^O_{\text{closed}}(z) } V_\phi(\nu)\rangle_{T^2} \\ 
  &  \underset{\substack{z= \nu-\epsilon \\ \epsilon \to 0}}{\sim}  &
  \!ig_s^4 
 \int \!\!d^2\!\tau\, d^2\!\epsilon\, d^2\!\nu \,
 {e^{-4\pi N{\text{Im}(\nu)^2 \ov \text{Im}(\tau)}}\ov (\text{Im}(\tau))^{{d+1 \ov 2}}}
   {T(d_c, d, \tau, \bar\tau) \ov |\eta(\tau)|^{2(D-2)}}
 \!\!\left|{\theta_{1}(\nu,\tau) \ov \theta'_{1}(0,\tau)} \right|^{4N}
 |\epsilon|^{-4q^2}\!\!\!\!\!\!
 \nonumber \\
 & & ~~~~~~ \times
 \mathcal{P}^X_\phi(W,\Omega,\partial_\nu \Omega,..,
 \tilde\Omega,\partial_{\bar\nu} \tilde\Omega,..,\vec q B,
 \vec q^{\, 2}\partial_\epsilon B,...\vec q \tilde B,
 \vec q^{\, 2}\partial_{\bar\epsilon} \tilde B,..)\,\nonumber \\
 & & ~~~~~~ \times \chi(\nu, \bar\nu, \tau, \bar\tau)  \nonumber 
 \ee
where
 \be
 \Omega &= &\partial_\nu^2 \ln(e^{-{\pi\text{Im}(\nu)^2 \ov \text{Im}(\tau)}}\theta_{1}(\nu,\tau))
  \\
 W & = & {2\pi \ov \text{Im}(\tau)} \\
 B & \underset{\epsilon \to 0}{=} &
 \partial_\epsilon \ln\left(e^{-{\pi \text{Im}(\epsilon)^2 \ov \text{Im}(\tau)}} \theta_1(\epsilon, \tau)\right). 
 \ee
$\mathcal{P}_\phi$ is a polynomial in  
$W\, \Omega,\, \partial_\nu \Omega$, higher derivatives of $\Omega$,
the anti-holomorfic $\tilde \Omega$ and its
derivatives, $B, \tilde B$ and their higher derivatives, such that:
 \beq
 \mathcal{P}^X_\phi(W,\Omega,\ldots, \tilde \Omega,\ldots,\vec q
 B,\vec q \tilde B, \ldots) 
   =
   \mathcal{P}_\phi(W,\Omega,\ldots,\tilde\Omega,\ldots)+
   \mathcal{P}'_\phi(\vec q B, \,\vec q \tilde B, \ldots)
 \eeq
and $\chi(\nu, \bar\nu, \tau, \bar\tau)$ is the fermionic part of the
amplitude; 
the bosonic string case
is recovered by substituting 1 to it. 
The contribution from the compactified dimensions is given by
$T(d_c, d, \tau, \bar\tau)$, which, assuming  for simplicity 
compactification on a torus, reads:
 \beq
 T(d_c, d, \tau, \bar\tau) = \prod_{i=1}^{d_c-d}\frac{1}{R_i}
  e^{\sum_{n_i,w_i}i\tau\left({n^i\ov R^i}+{w^iR^i\ov 4}\right)^2-i\bar\tau\left({n^i\ov R^i}-{w^iR^i\ov 4}\right)^2},
 \eeq
where $d_c$ is the critical (spatial) dimension, that is 25 for the
bosonic theory and 9 for the superstring.

Considering the form of formula (\ref{massshiftfixingNR}), together
with the diagrammatic representation in figure \ref{massshiftfixedNRfigure}, we
propose to identify the amplitude (\ref{oneloopfixedRN}), in the limit
$z\to\nu$, with:
 \beq \label{squaredmasshisftfixedNR}
 A^R_{T^2}=  
 \langle\phi|\widehat{\Delta M}^2\,\delta(R^2-\hat R^2)|\phi\rangle
 \eeq
and from this, by tracing over states at fixed squared mass and
appropriately normalizing, to arrive at (\ref{masshisftfixedNR}).

The form of $\mathcal{P}'_\phi$ depends on the various vertex operator
$V_\phi$ and it is very hard to say something in quantitative details
about it. We content ourselves to stress the presence of the term
$|\epsilon|^{-4\vec q^{~2}}$ in the amplitude integrand. Indeed, let us for a
moment neglect $\mathcal{P}'_\phi$, we find:
 \be \label{constrainingterm}
 A^R_{T^2} & \sim &
 A_{T^2}
  \int_{|\epsilon|<1} d^2\!\epsilon \,|\epsilon|^{-\vec q^{~2}}\nonumber \\
 & \sim & 
  -{A_{T^2}\ov 2} {1 \ov \vec q^{~2} -1} 
 \ee
where $A_{T^2}$ is the one-loop amplitude with two insertions of the
vertex operator $V_\phi$ (giving the mass-renormalization for the
state $|\phi\rangle$):
 \beq
 A_{T^2}= g_c^4\iint \langle V_\phi(0)V_\phi(\nu) \rangle_{T^2}.
 \eeq

If we Fourier transform the last factor and take the real part (we are
interested in the mass-shift, not in the decay), we find
something of the form
($J_n, Y_n$ are Bessel functions, $a, b$ real functions of $N$):
 \beq
 -A_{T^2}\int d\vec q\, e^{i \vec q \cdot \vec x}{1 \ov \vec q^{~2} -1} \sim 
 -{aJ_{{d-2 \ov 2}}(|x|)+bY_{{d-2 \ov 2}}(|x|) \ov |x|^{{d-2 \ov 2}}}
 \eeq
As a mass correction in the formulas for the entropy, this term would
favour  small 
$|x|$ (or penalize it, depending on the signs of $a$ and $b$). Note
also that, while the rest of the  
dependence on $q$ in (\ref{oneloopfixedRN}), coming from
$\mathcal{P}'_\phi$, depends strongly on 
the specific state $|\phi\rangle$ considered, this contribution will
be always present, and therefore also recovered in the averaging over
the states of an ensemble.

Our result is clearly incomplete, since we have neglected contributions 
that are not necessarily negligible in the limit $z \to \nu$. In any case it 
shows how terms that favour states with small average size
$R=|x|$ can possibly arise. 
The important ingredient is the extra modulus $z$ for the torus with 
punctures that we need to integrate over: this leads to the terms we are
interested in. 
Of course, the overall sign of the amplitude is extremely important for 
favouring or penalizing more compact string states.
Unfortunately, due to the lack of knowledge about vertex operators for
heavy massive 
perturbative strings, it has been impossible to estimate the full
result and we leave it for future research.

\subsection{Loop corrections for BPS states}

For non-BPS states carrying Neveu-Schwarz charges, the consideration of the
previous sections apply, once taken account of the fact that the
holomorfic and the anti-holomorfic exponential part of their vertex
operators are now different\footnote{Remember that our closed string
  states are extended only in the uncompactified dimensions, that is
  all their dependence on the compactified ones is in the exponential
  part of their vertex operator.}.

The question of corrections for the BPS states is particularly
subtle. We know that the counting of these states, 
and therefore their
entropy,  for fixed mass and
charge do not receive corrections due to the vanishing of their
two-points torus amplitude. Nevertheless, we have shown in (\ref{sizeBPS})
that their average size at zero coupling is larger than the string
scale, and therefore of the Schwarschild radius of the correspondent
black hole at the transition point.

Which is, then, the nature of the corrections that would favour more
compact states? The procedure described above fails for BPS states,
because of the vanishing of the torus two-point function $A_{T^2}$ in
(\ref{constrainingterm}). In particular, for the states described in
section \ref{BPSentropyfree}, the vanishing of $A_{T^2}$ is due to the
sum over spin structure with spin-statistic signs
for the holomorfic factor of the fermionic
contribution.
This implies that we cannot naively take the
OPE as in (\ref{massshiftfixingNR}), but we must {\em first} sum over
the spin structure. With the insertion  
of two vertex operators of the kind given in (\ref{gravvertex}), the
amplitude indeed does not vanish. 
On the other side, we cannot propose any more what we get as a self-energy
diagram with the insertion of a delta function, not even considering
certain limits, as done in the previous section.

Let us see why this is the case, trying to get as close as possible to
(\ref{massshiftfixingNR}).
The BPS states in section \ref{BPSentropyfree} are
represented by the vertex operators:
 \be \label{NSBPSoperator}
 V_\phi(z, \bar z)= \zeta \cdot \psi e^{i p_L\cdot X_L}e^{-i\varphi} 
  \tilde V_\phi(\bar z) & \text{in the Neveu-Schwarz sector} 
 \ee
\vspace{-0.8cm}
 \be
 V_\phi(z, \bar z)= u_\alpha S^\alpha e^{i p_L\cdot X_L}
  e^{-{i \ov 2}\varphi} \tilde V_\phi(\bar z) 
  & \text{in the Ramond sector} 
 \ee
where:
 \be
 \tilde V_\phi(\bar z) & 
 \text{is the anti-holomorphic part of the vertex operator}
 \nonumber \\
 p_L \equiv (p, \vec Q_L) & \text{$p$ is the momentum in the
 $d+1$ extended dimensions} \nonumber \\
 S^\alpha & \text{is the ground state spin field} \nonumber \\
 e^{-i\varphi},   e^{-{i \ov 2}\varphi} & \text{are the bosonized ground
 state operators for the superghost\footnotemark} \nonumber
 \ee
\footnotetext{In order to cancel the
   superghost charge anomaly, the vertex operator for the second
   insertion in the two-point function will have holomorphic part 
 in the (-1)
 and ($-{3 \ov 2}$) picture, respectively for the Neveu-Schwarz and
 Ramond sector.}

Summing over the spin structures and {\em then} trying to take a
limit in order to reproduce at best (\ref{massshiftfixingNR}), we
obtain\footnote{Formula (\ref{massshiftfixingNRBPS}) has been obtained
  for the case with vertex operator (\ref{NSBPSoperator}).}:
 \begin{multline} \label{massshiftfixingNRBPS}
 \langle V_{\phi'}(p', 0) V(k', z) V(k, \xi) V_\phi(p, \nu)\rangle_{T^2} 
 \underset{\substack{q^2 \to 0 \\ \xi = z+|\eta| \\ |\eta|\to 0}}{\sim} \\  
 ~~~~~~~~~~~~~~{1 \ov \vec q^{\,2}}\,
 \langle e^{i p_L\cdot X_L(0)}\tilde V_{\phi'}(p', 0)\,e^{i\vec q\cdot
   \hat{\vec{X}}^O_{L,\text{closed}}(z, \bar z)}\,|z|^{4k'\cdot \hat p}\,e^{i p_L\cdot X_L(\nu)}\tilde V_\phi(p,\bar \nu)\rangle_{T^2}.
 \end{multline}
Looking at (\ref{massshiftfixingNRBPS}) we see that in any case we do
not find a correlator such as (\ref{massshiftfixingNR}), but instead,
in the holomorphic factor, we do not have any more the (holomorphic
part of the) vertex operator for the BPS state. As we said, it is therefore not
possible to identify the amplitude $A^R_{T^2}$, obtained 
integrating over 
the correlator (\ref{massshiftfixingNRBPS}), with the formula
 \beq 
 A^R_{T^2}=  
 \langle\phi|\widehat{\Delta M}^2\, \delta(R^2-\hat R^2)|\phi\rangle
 \eeq 

In any case, we can try to study this amplitude as a correction to the
form factor, 
and so we find a term, by letting $z= \nu-\epsilon, \, \epsilon \to 0$:
 \be
 A^R_{T^²} 
  & 
  \underset{\substack{z= \nu-\epsilon\\ \epsilon \to 0}}{\sim} &
  ig_c^4
 \int \!\!d^2\!\tau\, d^2\!\epsilon\, d^2\!\nu \,
 {e^{-4\pi N{\text{Im}(\nu)^2 \ov \text{Im}(\tau)}}\ov (\text{Im}(\tau))^{{d+1 \ov 2}}}
   {T(d_c, d, \tau, \bar\tau) \ov |\eta(\tau)|^{2(D-2)}}
 \!\!\left|{\theta_{1}(\nu,\tau) \ov \theta'_{1}(0,\tau)} \right|^{4N}
 |\epsilon|^{-4q^2}\!\!\!\!\!\!
 \nonumber \\
 & & ~~~~~~ \times
 \mathcal{P}^X_\phi(\tilde\Omega,\partial_{\bar\nu} \tilde\Omega,..,\vec q \tilde B, \vec q^{\, 2}\partial_\epsilon \tilde B,..)\,\nonumber \\
 & & ~~~~~~ \times \chi(\bar\nu, \bar\tau)   
 \ee
which again favours compact sizes for the string, as we argued in the
previous section. 

\section{Conclusions}

This work has dealt with two principal topics: the degeneracy of
perturbative closed superstring states depending on their mass,
charges and average size 
and the dynamics of such states under self-interactions.

Our principal result is the
formula for the entropy of string states at fixed mass, charge and size
for the free string case ($g_c=0$). 
Its derivation has required the proper definitions of
suitable operators and density matrices for the ensemble under
investigation. The computations are well-defined within
Superstring Theory, since we have obtained our results starting from
well-defined string amplitudes. The key-point has
been the property of factorization of the amplitudes. Indeed, string
theory is defined only when the external legs of an amplitude are
on-shell; however using factorization, we can operate on the momenta
flowing in the 
internal lines, which allows analytical
continuation. 

The entropy formula for free string states generically
suffers from corrections, due to the
self-interaction of the string at non-zero coupling. 
We have investigated the one-loop corrections for states
constrained in size as well as in mass. Unfortunately the complexity
of the computation has hindered to obtain a full result, but we have
shown how states with compact size can be possibly favoured by the
self-interaction of the string. Our analysis has again relied on the
factorization properties of well-defined string amplitudes.

\section{Acknowledgements}

I would like to thank Ulf Danielsson for his useful remarks in
discussing the results of this work. I am
grateful also for the conversations I had with other colleagues, such as
 Valentina Giangreco, Niklas Johansson, Thomas Klose, Konstantin Zarembo.

This work was supported by the EU Marie Curie Training Site contract:
MRTN-CT-2004-512194.



\end{document}